\documentclass[journal, onecolumn, 11 pt]{IEEEtran}
\newlength{\tocsep}
\usepackage{makeidx}

\usepackage{epsfig,float,multirow,footmisc,rotating}
\usepackage{graphics}
\usepackage{graphicx}
\usepackage{color}
\usepackage{placeins}

\usepackage[utf8]{inputenc}
\usepackage{verbatim}
\usepackage{url}
\usepackage{array}
\DeclareGraphicsExtensions{.jpg, .png, .pdf, .tiff}

\usepackage{wrapfig}
\usepackage{sidecap}
\usepackage{multicol}

\usepackage[labelfont=bf]{caption}
\usepackage{subcaption}
\usepackage{multimedia}
\usepackage[]{units}
\usepackage{amsmath,amsfonts,amssymb}
\usepackage[normalem]{ulem}
\usepackage{environ}
\usepackage{afterpage}

\definecolor{darkgreen}{rgb}{0.0, 0.2, 0.13}
\definecolor{darkolivegreen}{rgb}{0.33, 0.42, 0.18}

\newcommand{\eg}{\textit{e.g.}\ }
\newcommand{\ie}{\textit{i.e.}\ }

\begin{document}

\title{Evolutionary optimisation of neural network models for fish collective behaviours in mixed groups of robots and zebrafish}

\author{
Leo Cazenille$^{1,2}$,
Nicolas Bredeche$^{2}$,
Jos\'{e} Halloy$^{1}$\\

{1} Univ Paris Diderot, Sorbonne Paris Cit\'e, LIED, UMR 8236, 75013, Paris, France
\\{2} Sorbonne Universit\'e, CNRS, ISIR, F-75005 Paris, France
}

\maketitle

\begin{abstract}
Animal and robot social interactions are interesting both for ethological studies and robotics. On the one hand, the robots can be tools and models to analyse animal collective behaviours, on the other hand the robots and their artificial intelligence are directly confronted  and compared to the natural animal collective intelligence. The first step is to design robots and their behavioural controllers that are capable of socially interact with animals. Designing such behavioural bio-mimetic controllers remains an important challenge as they have to reproduce the animal behaviours and have to be calibrated on experimental data. Most animal collective behavioural models are designed by modellers based on experimental data. This process is long and costly because it is difficult to identify the relevant behavioural features that are then used as \textit{a priori} knowledge in model building. Here, we want to model the fish individual and collective behaviours in order to develop robot controllers. We explore the use of optimised black-box models based on artificial neural networks (ANN) to model fish behaviours. While the ANN may not be biomimetic but rather bio-inspired, they can be used to link perception to motor responses. These models are designed to be implementable as robot controllers to form mixed-groups of fish and robots, using few \textit{a priori} knowledge of the fish behaviours. We present a methodology with multilayer perceptron or echo state networks that are optimised through evolutionary algorithms to model accurately the fish individual and collective behaviours in a bounded rectangular arena.  We assess the biomimetism of the generated models and compare them to the fish experimental behaviours.
\end{abstract}


\section{Introduction}
Autonomous, biomimetic robots can serve as tools in animal behavioural studies. Robots are used in ethology and behavioural studies to untangle the multimodal modes of interactions and communication between animals~\cite{mondada2013general}. When they are socially integrated in a group of animals, they are capable of sending calibrated stimuli to test the animal responses in a social context~\cite{halloy2007social}. Moreover, animal and autonomous robot interactions represent an interesting challenge for robotics. Confronting robots to animals is a difficult task because specific behavioural models have to be designed and the robots have to be socially accepted by the animals. The robots have to engage in social behaviour and convince somehow the animal that they can be social companions. In this context, the capabilities of the robots and their intelligence are put in harsh conditions and often demonstrate the huge gap that still exists between autonomous robots and animals not only considering motion and coping with the environment but also in terms of intelligence. It is a direct comparison of artificial and natural collective intelligence.
Moreover, the design of such social robots is challenging as it involves both a luring capability including appropriate robot behaviours, and the social acceptation of the robots by the animals. We have shown that the social integration of robots into groups of fish can be improved by refining the behavioural models used to build their controllers~\cite{cazenille2017acceptation}. The models have also to be calibrated to replicate accurately the animal collective behaviours in complex environments~\cite{cazenille2017acceptation}.

Research on animal and robot interactions need also bio-mimetic formal models as behavioural controllers of the robots if the robots have to behave as congeners~\cite{Bonnet2016IJARS,bonnet2017cats}. Robots controllers have to deal with a whole range of behaviours to allow them to take into account not only the other individuals but also the environment and in particular the walls \cite{cazenille2017acceptation,cazenille2017automated}. However, most of biological collective behaviour models deal only with one sub-part at a time of fish behaviours in unbounded environments. Controllers based on neural networks, such as multilayer perceptron (MLP)~\cite{king1989neural} or echo state networks (ESN)~\cite{jaeger2007echo} have the advantage to be easier to implement and could deal with a larger range of events.

\subsection*{Objectives}
We aim at building models that generate accurately zebrafish trajectories of one individual within a small group of 5 agents. The trajectories are the result of social interactions in a bounded environment. Zebrafish are a classic animal model in the research fields of genetics and neurosciences of individual and collective behaviours. Building models that correctly reproduce the individual trajectories of fish within a group is still an open question~\cite{herbert2015turing}. We explore MLP and ESN models, optimised by evolutionary computation, to generate individual trajectories. MLP and ESN are black-box models that need few \textit{a priori} information provided by the modeller. They are optimised on the experimental data and as such represent a model of the complex experimental collective trajectories. However, they are difficult to calibrate on the zebrafish experimental data due to the complexity of the fish trajectories.
Here, we consider the design and calibration by evolutionary computation of neural network models, MLP and ESN, that can become robot controllers. We test two evolutionary optimisation methods, CMA-ES~\cite{auger2005restart} and NSGA-III~\cite{yuan2014improved} and show that the latter gives better results. We show that such MLP and ESN behavioural models could be useful in animal robot interactions and could make the robots accepted by the animals by reproducing their behaviours and trajectories as in~\cite{cazenille2017acceptation}.

\section{Materials and Methods} \label{sec:methods}

\subsection{Experimental set-up} \label{sec:setup}
We use the same experimental procedure, fish handling, and set-up as in~\cite{cazenille2016automated,bonnet2017cats,seguret2017loose,cazenille2017acceptation,collignon2017collective,bonnet2018closed}. The experimental arena is a square white plexiglass aquarium of $1000\times1000\times100$~mm. An overhead camera captures frames at 15 FPS, with a $500\times500$px resolution, that are then tracked to find the fish positions. We use 10 groups of 5 adults wild-type AB zebrafish (\textit{Danio rerio}) in 10 trials lasting each one for 30-minutes as in~\cite{cazenille2016automated,bonnet2017cats,seguret2017loose,cazenille2017acceptation,collignon2017collective,bonnet2018closed}.
The experiments performed in this study were conducted under the authorisation of the Buffon Ethical Committee (registered to the French National Ethical Committee for Animal Experiments \#40) after submission to the French state ethical board for animal experiments.

\begin{figure*}[h]
\begin{center}
\includegraphics[width=0.85\textwidth]{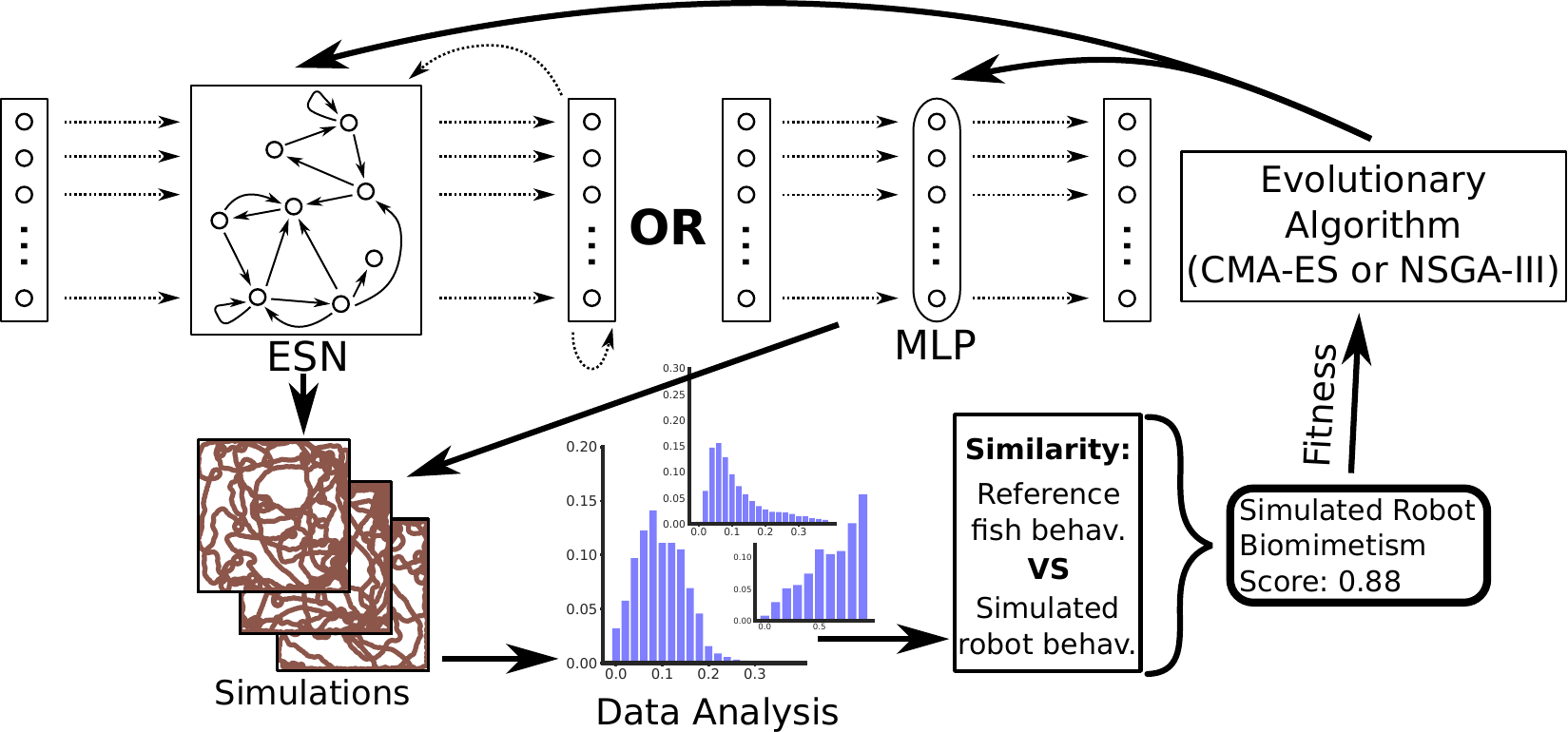}
\caption{Methodology workflow. An evolutionary algorithm is used to evolve the weight of a MLP (1 hidden layer, 100 neurons) or an ESN (100 reservoir neurons) neural networks that serves as the controller of a simulated robot interacting with 4 fish described by the experimental data. Only the connections represented by dotted arrows are evolved (for MLP: all connections; for ESN: connections from inputs to reservoir, from reservoir to outputs and from outputs to outputs and to reservoir). The fitness function is computed through data-analysis of these simulations and represent the biomimetism metric of the simulated robot behaviour compared to the behaviour exhibited by real fish in experiments. Two evolutionary algorithms are tested: CMA-ES (mono-objective) and NSGA-III (multi-objective).}
\label{fig:workflow}
\end{center}
\end{figure*}

\subsection{Artificial neural network model} \label{sec:model}
Black-box models, like artificial neural networks (ANN), can be used to model phenomena with few \textit{a priori} information. Although they are not used yet to model fish collective behaviours based on experimental data, here we show that they are relevant to model zebrafish collective behaviour.
We propose a methodology (Fig.~\ref{fig:workflow}) where either a multilayer perceptron (MLP)~\cite{king1989neural} artificial neural network, or an echo state network (ESN)~\cite{jaeger2007echo}, is calibrated through the use of evolutionary algorithms to model the behaviour of a simulated fish in a group of 5 individuals. The 4 other individuals are described by the experimental data obtained with 10 different groups of 5 fish for trials lasting 30 minutes.

MLP are a type of feedforward artificial neural networks that are very popular in artificial intelligence to solve a large variety of real-world problems~\cite{norgaard2000neural}. Their capability to universally approximate functions~\cite{cybenko1989approximation} makes them suitable to model control and robotic problems~\cite{norgaard2000neural}. We consider MLP with only one hidden layer of $100$ neurons (using a hyperbolic tangent function as activation function).

ESN are recurrent neural networks often used to model temporal processes, like time-series, or robot control tasks~\cite{polydoros2015advantages}. They are sufficiently expressive to model complex non-linear temporal problems, that non-recurrent MLP cannot model.

For the considered focal agent, the neural network model takes the following parameters as input:
    (i) the \textit{direction vector} (angle and distance) from the focal agent towards each other agent;
    (ii) the \textit{angular distance} between the focal agent direction and each other agent direction (alignment measure);
    (iii) the \textit{direction vector} (angle and distance) from the focal agent towards the nearest wall;
    (iv) the \textit{instant linear speed} of the focal agent at the current time-step, and at the previous time-step;
    (v) the \textit{instant angular speed} of the focal agent at the current time-step, and at the previous time-step.
This set of inputs is typically used in multi-agent modelling of animal collective behaviour~\cite{deutsch2012collective,sumpter2012modelling}. As a first step, we consider that it is sufficient to model fish behaviour with neural networks.

The neural network has two outputs corresponding to the change in linear and angular speeds to apply from the current time-step to the next time-step. Here, we limit our approach to modelling fish trajectories resulting from social interactions in a homogeneous environment but bounded by walls. Very few models of fish collective behaviours take into account the presence of walls~\cite{collignon2016stochastic,calovi2018disentangling}.

\subsection{Data analysis} \label{sec:dataAnalysis}
For each trial, $e$, and simulations, we compute several behavioural metrics using the tracked positions of agents: (i) the distribution of \textit{inter-individual distances} between agents ($D_e$); (ii) the distributions of \textit{instant linear speeds} ($L_e$); (iii) the distributions of \textit{instant angular speeds} ($A_e$); (iv) the distribution of \textit{polarisation} of the agents in the group ($P_e$) and (v) the distribution of \textit{distances of agents to their nearest wall} ($W_e$).
The polarisation of an agent group measures how aligned the agents in a group are, and is defined as the absolute value of the mean agent heading: $P = \frac{1}{N} \bigl\lvert  \sum^{N}_{i=1} u_i \bigr\rvert$ where $u_i$ is the unit direction of agent $i$ and $N=5$ is the number of agents~\cite{tunstrom2013collective}.

We define a similarity measure (ranging from $0.0$ to $1.0$) to measure the biomimetism of the simulated robot behaviour by comparing the behaviour of the group of agents in simulations where the robot is present (experiment $e_r$: four fish and one robot) to the behaviour of the experimental fish groups (experiment $e_c$: five fish):
\begin{equation}
S(e_r, e_c) = \sqrt[5]{I(D_{e_r}, D_{e_c}) I(L_{e_r}, W_{e_c}) I(A_{e_r}, O_{e_c}) I(P_{e_r}, T_{e_c}) I(W_{e_r}, T_{e_c})}
\end{equation}
The function $I(X, Y)$ is defined as such: $I(X, Y) = 1 - H(X, Y)$.
The $H(X, Y)$ function is the Hellinger distance between two histograms~\cite{deza2006dictionary}. It is defined as: $H(X, Y) = \frac{1}{\sqrt{2}} \sqrt{ \sum_{i=1}^{d} (\sqrt{X_i} - \sqrt{Y_i}  )^2 }$ where $X_i$ and $Y_i$ are the bin frequencies.

This score measures the social acceptation of the robot by the fish, as defined in~\cite{cazenille2017acceptation,cazenille2017automated}. Compared to the similarity measure defined in these articles, we added a measure of the polarisation of the agents. This was motivated by the tendency of our evolved neural models, without a polarisation factor, to generate agents with unnatural looping behaviour to catch up with the group.

\subsection{Optimisation} \label{sec:optim}
We calibrate the ANN models presented here to match as close as possible the behaviour of one fish in a group of 5 individuals in 30-minute simulations (at $15$ time-steps per seconds, \ie $27000$ steps per simulation). This is achieved by optimising the connection weights of the ANN through evolutionary computation that iteratively perform global optimisation (inspired by biological evolution) on a defined fitness function so as to find its maxima~\cite{salimans2017evolution,jiang2008supervised}. 

We consider two optimisation methods (as in~\cite{cazenille2017automated}), for MLP and ESN networks.
In the \textbf{Sim-MonoObj-MLP} case, we use the CMA-ES~\cite{auger2005restart} mono-objective evolutionary algorithm to optimise an MLP, with the task of maximising the $S_(e_1, e_2)$ function.
In the \textbf{Sim-MultiObj-MLP} and \textbf{Sim-MultiObj-ESN} cases, we use the NSGA-III~\cite{yuan2014improved} multi-objective algorithm with three objectives to maximise. The first objective is a performance objective corresponding to the $S_(e_1, e_2)$ function. We also consider two other objectives used to guide the evolutionary process: one that promotes genotypic diversity~\cite{mouret2012encouraging} (defined by the mean euclidean distance of the genome of an individual to the genomes of the other individuals of the current population), the other encouraging behavioural diversity (defined by the euclidean distance between the $D_{e}$, $L_{e}$, $A_{e}$, $P_{e}$ and $W_{e}$ scores of an individual). The NSGA-III algorithm was used with a $0.80\%$ probability of crossovers and a $0.20\%$ probability of mutations (we also tested this algorithm with only mutations and obtained similar results).
The NSGA-III algorithm~\cite{yuan2014improved} is considered instead of the NSGA-II algorithm~\cite{deb2002fast} employed in~\cite{cazenille2017automated} because it is known to converge faster than NSGA-II on problems with more than two objectives~\cite{ishibuchi2016performance}.

In both methods, we use populations of 60 individuals and 300 generations. Each case is repeated in 10 different trials.
We use a NSGA-III implementation based on the DEAP python library~\cite{fortin2012deap}.

\begin{figure}[h]
\centering
\includegraphics[width=0.99\textwidth]{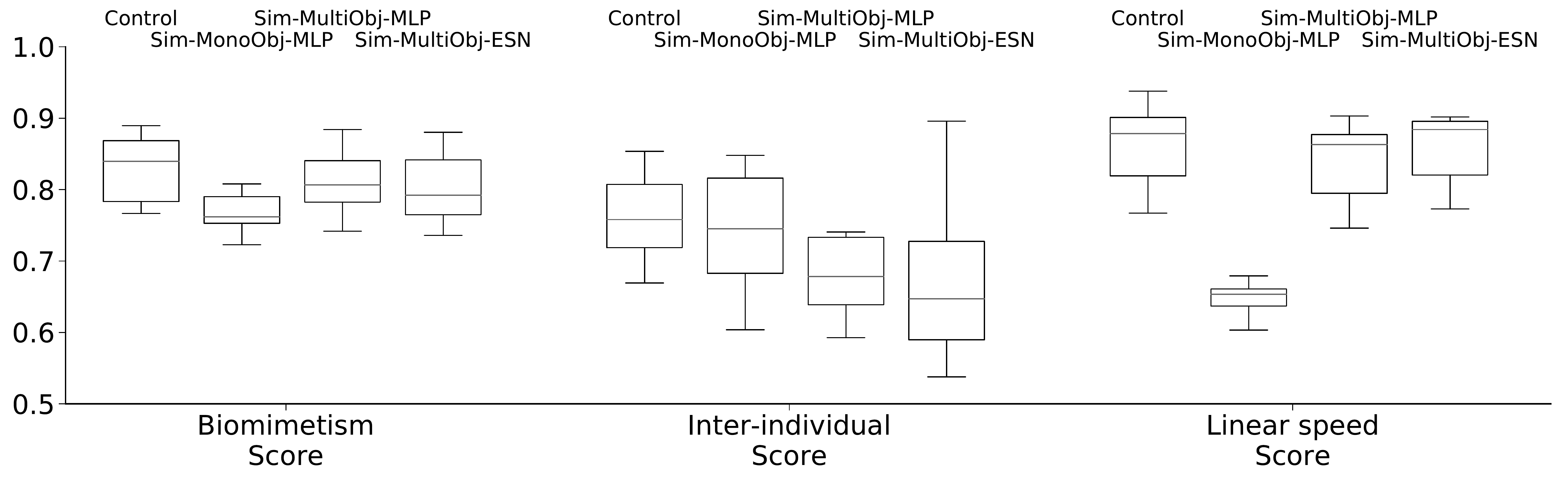}
\includegraphics[width=0.99\textwidth]{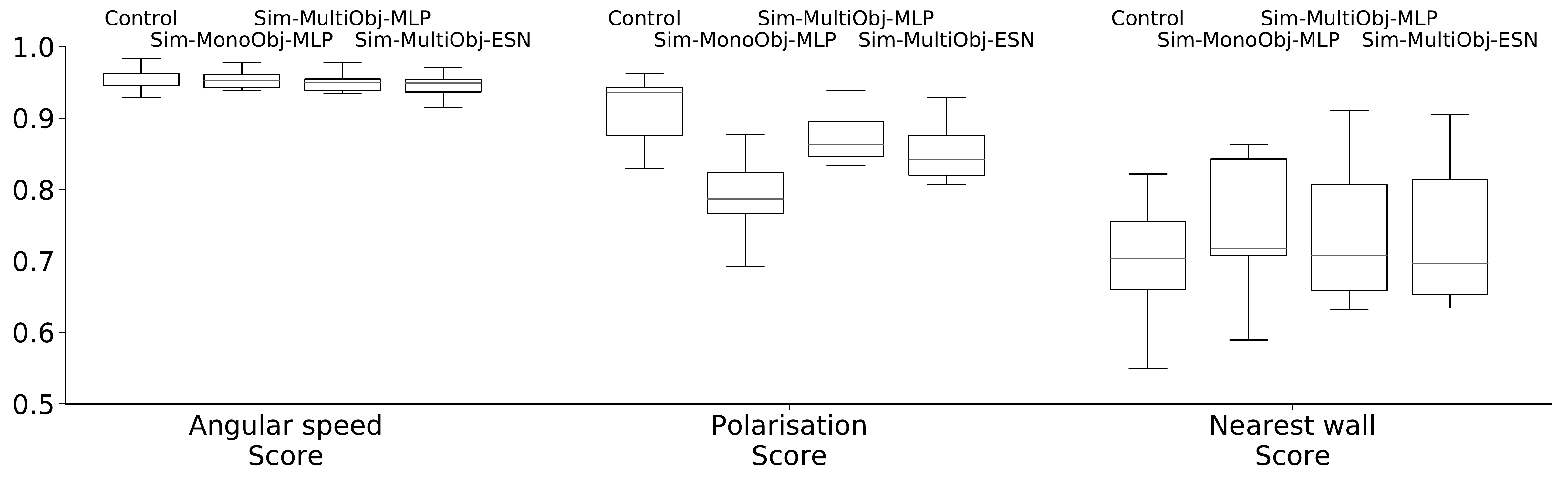}
\caption{Similarity scores between the behaviour of the experimental fish groups (control) and the behaviour of the best-performing simulated individuals of the MLP models optimised by CMA-ES or NSGA-III.  Results are obtained over 10 different trials (experiments for fish-only groups, and simulations for NN models). We consider five behavioural features to characterise exhibited behaviours. \textbf{Inter-individual distances} corresponds to the similarity in distribution of inter-individual distances between all agents and measures the capabilities of the agents to aggregate. \textbf{Linear and Angular speeds distributions} correspond to the distributions of linear and angular speeds of the agents. \textbf{Polarisation} measures how aligned the agents are in the group. \textbf{Distances to nearest wall} corresponds to the similarity in distribution of agent distance to their nearest wall, and assess their capability to follow the walls. The \textbf{Biomimetic score} corresponds to the geometric mean of the other scores.}
\label{fig:scores}
\end{figure}

\begin{figure}[h]
\begin{center}
\includegraphics[width=0.90\textwidth]{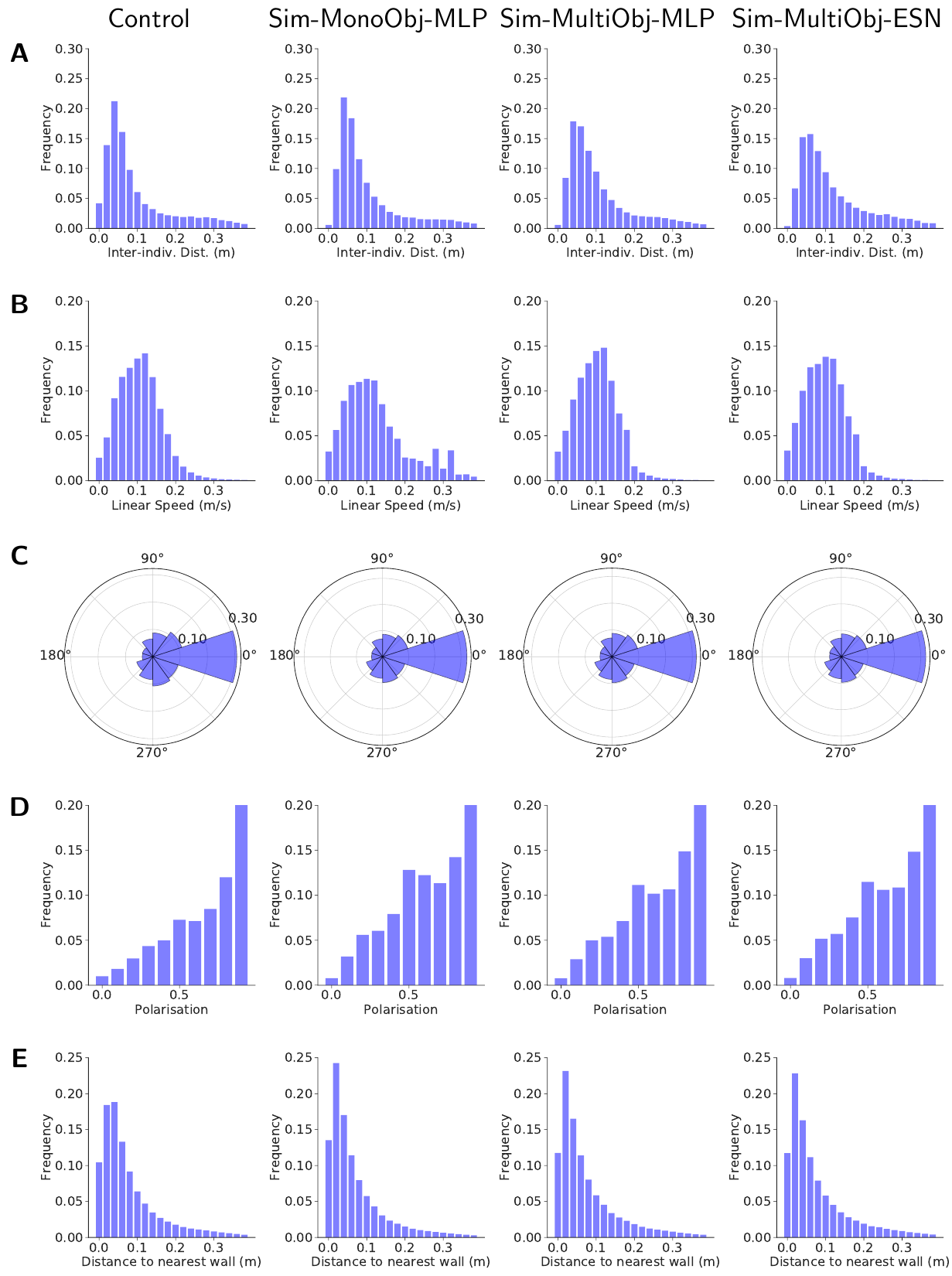}
\caption{Comparison between 30-minutes trials involving 5 fish (control, biological data) and simulations involving 4 fish and 1 robot, over 10 trials and across 5 behavioural features: inter-individual distances (\textbf{A}), linear (\textbf{B}) and angular (\textbf{C}) speeds distributions, polarisation (\textbf{D}), and distances to nearest wall (\textbf{E}).}
\label{fig:plotsHists}
\end{center}
\end{figure}

\section{Results} \label{sec:results}
We analyse the behaviour of one simulated robot in a group of 4 fish. The robots are driven by ANN (either MLP or ESN) evolved with CMA-ES (\textbf{Sim-MonoObj-MLP} case) or with NSGA-III (\textbf{Sim-MultiObj-MLP} and \textbf{Sim-MultiObj-ESN} cases) and compare it to the behaviour of fish-only groups (\textbf{Control} case). We only consider the best-evolved ANN controllers. In the simulations, the simulated robot does not influence the fish because the fish are described by their experimental data that is replayed.

\begin{figure}[h]
\begin{center}
\includegraphics[width=0.80\textwidth]{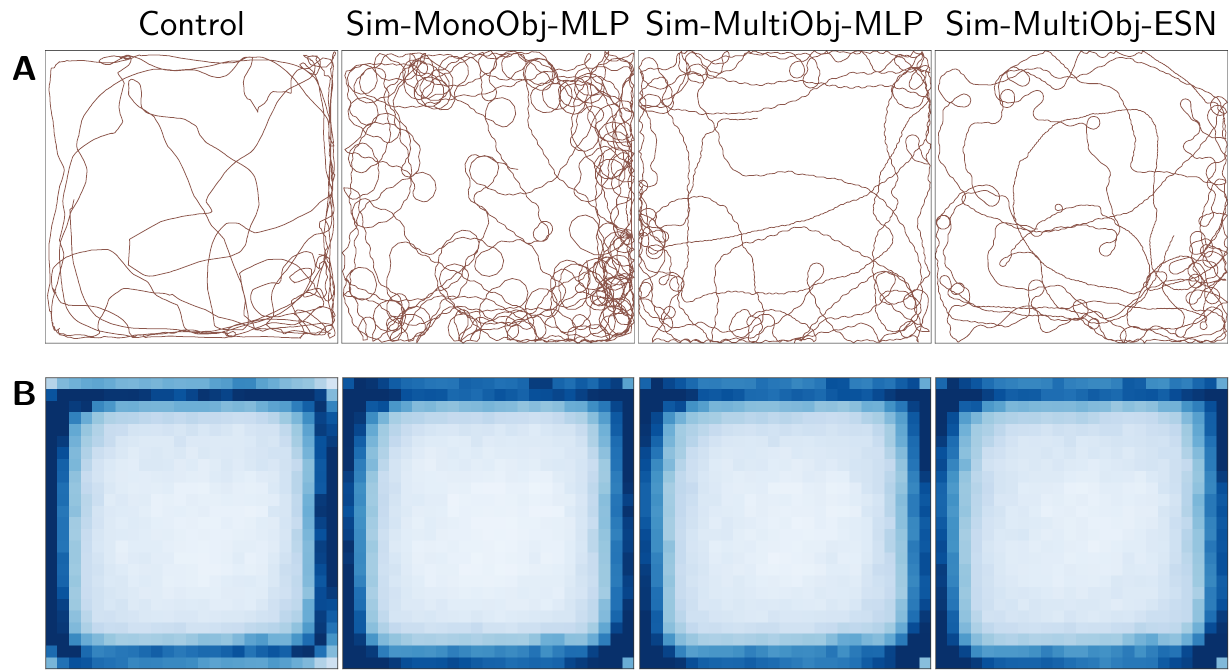}
\caption{Agent trajectories observed after 30-minute trials in a square ($1m$) aquarium, for the 4 considered cases: \textbf{Control} reference experimental fish data obtained as in~\cite{collignon2016stochastic,seguret2017loose}, \textbf{Sim-MonoObj-MLP} MLP optimised by CMA-ES, \textbf{Sim-MultiObj-MLP} MLP optimised by NSGA-III, \textbf{Sim-MultiObj-ESN} ESN optimised by NSGA-III. \textbf{A} Examples of an individual trajectory of one agent among the 5 making the group (fish or simulated robot) during 1-minute out of a 30-minute trial. \textbf{B} Presence probability density of agents in the arena.}
\label{fig:plotsTraces}
\end{center}
\end{figure}

Examples of agent trajectories obtained in the three tested cases are found in Fig.~\ref{fig:plotsTraces}A. In the \textbf{Sim-MonoObj-MLP} and \textbf{Sim-MultiObj-*} cases, they correspond to the trajectory of the simulated robot agent.
In both case, we can see that the robot follow the walls like the fish, and are often part of the fish group as natural fish do. However, the robot trajectories can incorporate patterns not found in the fish trajectories. For example, small circular loop are done when the robot performs an U-turn to catch up with the fish group. This is particularly present in the \textbf{Sim-MonoObj-MLP} case, and seldom appear in the \textbf{Sim-MultiObj-*} cases.

We compute the presence probability density of agents in the arena (Fig.~\ref{fig:plotsTraces}B): it shows that the robot tend to follow the walls as the fish do naturally.

For the three tested cases, we compute the statistics presented in Sec.~\ref{sec:dataAnalysis} (Fig.~\ref{fig:plotsHists}). The corresponding similarity scores are shown in Fig.~\ref{fig:scores}. The results of the \textbf{Control} case shows sustained aggregative and wall-following behaviours of the fish group. Fish also seldom pass through the centre of the arena, possibly in small short-lived sub-groups. There is group behavioural variability, especially on aggregative tendencies (measured by inter-individual distances), and wall-following behaviour (measured by the distance to the nearest wall), because each one of the 10 groups is composed of different fish \textit{i.e.} 50 fish in total.

The similarity scores of the \textbf{Sim-MultiObj-*} cases are often within the variance domain of the \textbf{Control} case, except for the inter-individual score. It suggests that groups incorporating the robot driven by an MLP evolved by NSGA-III exhibit relatively similar dynamics as a fish-only group, at least according to our proposed measures. However, it is still perfectible: the robot is sometimes at the tail of the group, possibly because of gap created between the robot and the fish group by small trajectories errors (\eg small loops shown in robot trajectories in Fig.~\ref{fig:plotsTraces}A).

The \textbf{Sim-MonoObj-MLP} case sacrifices biomimetism to focus mainly on group-following behaviour: this translated into a higher inter-individual score than in the \textbf{Sim-MultiObj-*} cases, and robot tend to follow closely the fish group. With \textbf{Sim-MonoObj-MLP}, the robot is going faster than the fish, and will fastly go back towards the centroid of the group if it is too far ahead of the group: this explains the large presence of loops in Fig.~\ref{fig:plotsTraces}A. The \textbf{Sim-MonoObj-MLP} does not take into account behavioural diversity like the \textbf{Sim-MultiObj-*}, but focus on the one that is easier to find (namely the group-following behaviour) and stays stuck in this local optimum.

There are few differences between the results of the \textbf{Sim-MultiObj-MLP} and the \textbf{Sim-MultiObj-ESN} cases, the latter showing often slightly lower scores than the former. However, the \textbf{Sim-MultiObj-ESN} displays a large variability of inter-individual scores, which could suggest that its expressivity could be sufficient to model agents with more biomimetic behaviours if the correct connection weights were found by the optimiser.

\section{Discussion and Conclusion} \label{sec:conclusion}
We evolved artificial neural networks (ANN) to model the behaviour of a single fish in a group of 5 individuals. This ANN controller was used to drive the behaviour of a robot agent in simulations to integrate the group of fish by exhibiting biomimetic behavioural capabilities. Our methodology is similar to the calibration methodology developed in~\cite{cazenille2017automated}, but employs artificial neural networks instead of an expert-designed behavioural model. Artificial neural networks are black-box models that require few \textit{a-priori} information about the target tasks.

We design a biomimetism score from behavioural measures to assess the biomimetism of robot behaviour. In particular, we measure the aggregative tendencies of the agents (inter-individual distances), their disposition to follow walls, to be aligned with the rest of the group (polarisation), and their distribution of linear and angular speeds.

However, finding ANN displaying behaviours of appropriate levels of biomimetism is a challenging issue, as fish behaviour is inherently multi-level (tail-beats as motor response vs individual trajectories vs collective dynamics), multi-modal (several kinds of behavioural patterns, and input/output sources), context-dependent (different behaviours depending on the spatial position and proximity to other agents) and stochastic (leading to individual and collectives choices and action selection)~\cite{collignon2016stochastic,sumpter2018using}. More specifically, fish dynamics involve trade-offs between social tendencies (aggregation, group formation), and response to the environment (wall-following, zone occupation); they also follow distinct movement patterns that allow them to move in a polarised group and react collectively to environmental and social cues.

We show that this artificial neural models can be optimised by using evolutionary algorithms, using the biomimetism score of robot behaviour as a fitness function. The best-performing evolved ANN controllers show competitive biomimetism scores compared to fish group behavioural variability. We demonstrate that taking into account genotypic and behavioural diversity in the optimisation process (through the use of the global multi-objective optimiser NSGA-III) improve the biomimetic scores of the evolved best-performing controllers.
The ANN models evolved through mono-objective optimisation tend to focus more on evolving a group-following behaviour rather than a biomimetic agent.

Our approach is still perfectible, in particular, we only evolve the behaviour of a single agent in a group, rather than all agents of the group. This choice was motivated by the large increase in difficulty in evolving ANN models for the entire group, which would also involve additional behavioural trade-offs: \eg individual free-will and autonomous dynamics, individuals leaving or re-joining the group. However, it also means that here the fish do not react to the robot in simulations because the fish behaviour is a replay of fish experimental trajectories recorded without robot.

Additionally, it may be possible to improve the performance (in term of biomimetism) of the multi-objective optimisation process by combining additional selection pressures as objectives (\ie not just genotypic and behavioural diversity)~\cite{doncieux2014beyond}. We already include behavioural and phenotypic diversities as selection pressures to guide the optimisation process; however, taking into account phenotypic diversity can bias the optimisation algorithm to explore rather than exploit, which can prevent some desired phenotypes to be considered by the optimisation algorithm. An alternative would be to use angular diversity instead~\cite{szubert2016reducing}.

This study shows that ANN are good candidates to model individual and collective fish behaviours, in particular in the context of social bio-hybrid systems composed of animals and robots. By evolutionary computation, they can be calibrated on experimental data. This approach requires less \textit{a priori} knowledge than equations or agent based modelling techniques. Although they are black box model, they could also produce interesting results from a biological point of view. Thus, ANN collective behaviour models can be an interesting approach to design animal and robot social interactions.

\section*{Acknowledgement}
{\small
This work was funded by EU-ICT project 'ASSISIbf', no 601074.}

\FloatBarrier

\end{document}